\documentclass[twocolumn,showpacs]{revtex4}

\usepackage{amsmath}

\usepackage{graphicx}

\begin{document}
\draft
\title{Surface superconducting states in a polycrystalline MgB$_{2}$
sample}

\author{Menachem I. Tsindlekht$^a$, Grigory I. Leviev$^a$, Valery M. Genkin$^a$,
 Israel Felner$^a$, Pavlo Mikheenko$^b$, and J. Stuart Abell$^b$ }
\affiliation{$^a$The Racah Institute of Physics, The Hebrew
University of Jerusalem, 91904 Jerusalem, Israel; $^b$School of
Engineering, Department of Metallurgy and Materials, The University
of Birmingham, Edgbaston, B15 2TT, Birmingham, UK }

\begin{abstract}

We report results of dc magnetic and ac linear low-frequency study
of a polycrystalline MgB$_2$ sample. AC susceptibility measurements
at low frequencies, performed under dc fields parallel to the sample
surface, provide a clear evidence for surface superconducting states
in MgB$_2$.
\end{abstract}

\date{\today}
\maketitle

Nucleation of superconducting (SC) phase in the surface sheath under a dc magnetic field
parallel to the surface was predicted by Saint-James and de Gennes more than 40 years
ago~\cite{PG}. This prediction was made in the frame of the one-band isotropic
Ginzburg-Landau (GL) model and experimental confirmations of this prediction were
described in various publications (see, for example,~\cite{ROLL}). The discovery of
superconductivity in two-band anisotropic MgB$_2$ raised a question about the existence
of surface superconducting states (SSS) in this material. Transport measurements under dc
fields indicate that the onset of the SC transition occurs above the second critical
field, $H_{c2}$ determined from specific heat data~\cite{LYARD,RYDH}.

In this work we present the experimental results of dc and ac
measurements on polycrystalline MgB$_{2}$ sample. It appears that
large losses with a maximum at certain temperature dependent dc
field and complete diamagnetic screening of an ac field are observed
in magnetic fields when the dc magnetic moment is very small. This
is clear evidence for the existence of SSS in this material.

The MgB$_2$ sample with $T_c=37.5$~K and $\Delta T_c=0.5$~K was prepared using the Hot
Isostatic Pressing method~\cite{PASHA1}. The sample dimensions are $10\times 3\times 1$
mm$^3$. DC magnetization curves were measured by a SQUID magnetometer. ac susceptibility
was measured by the pick-up coils method~\cite{SH} in the frequency range
$5\leq\omega/2\pi\leq 1065$~Hz. An ac "home-made" setup was adapted to a commercial SQUID
magnetometer and its block diagram was published elsewhere~\cite{LEV2}. The experiments
were carried out as follows. The sample was cooled down in zero magnetic field, then both
dc ($H_0$) and ac magnetic field with amplitude $h_0$ were applied and the amplitude and
phase of ac response was measured. Both $H_0$ and $h_0$ were parallel to the longest
sample axis.

\begin{figure}
     \begin{center}
    \leavevmode
 \includegraphics[width=0.9\linewidth]{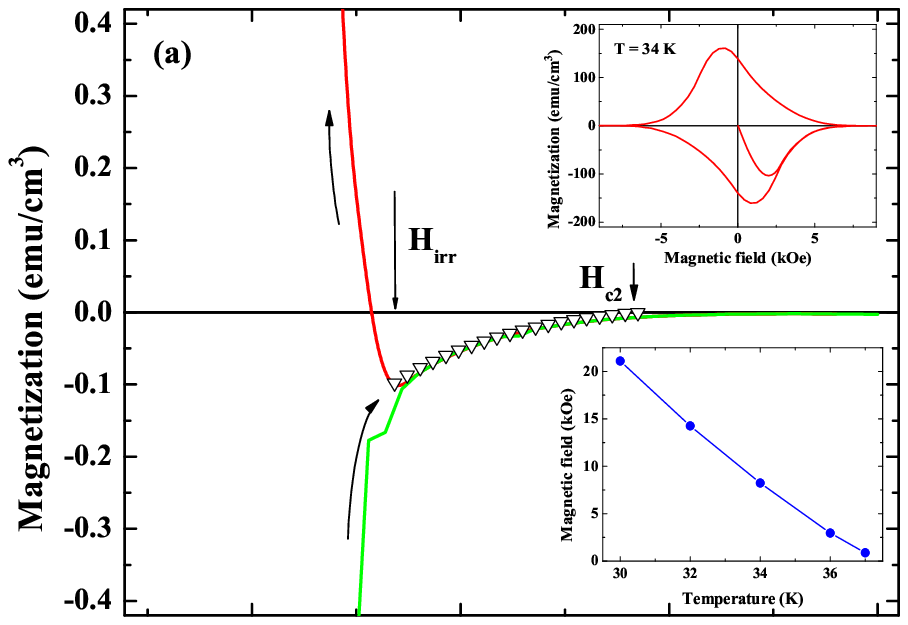}
\includegraphics[width=0.9\linewidth]{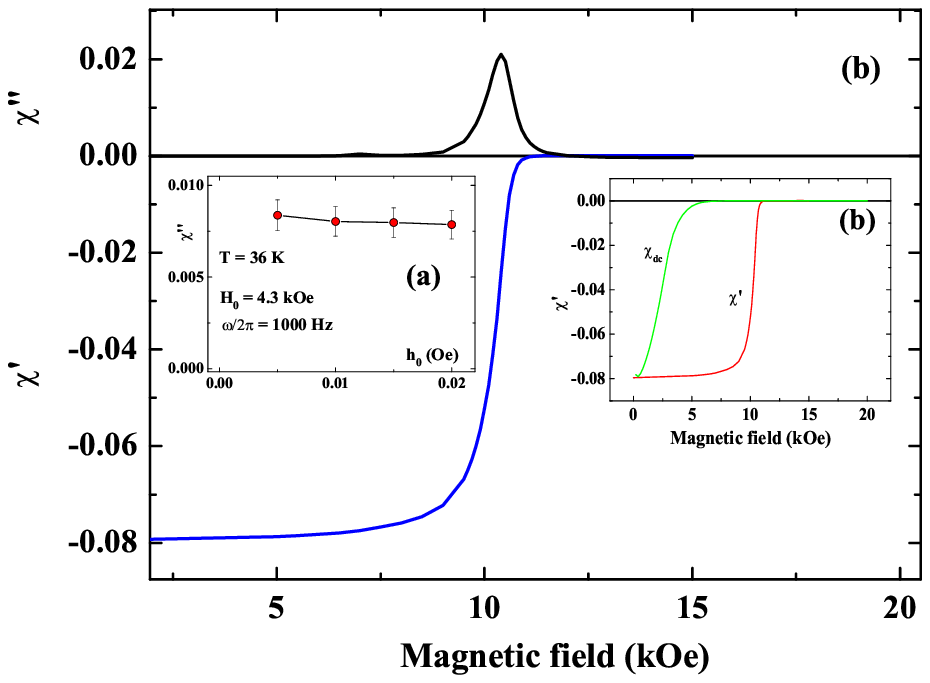}

       \caption{(Color online) (a) Magnetization curve, $M(H_0)$, in high magnetic fields.
      $\nabla$ - theoretical calculations for two-band model.
       Upper inset - magnetization loop after ZFC at $T=34$~K. Lower
       inset - $H_{irr}(T)$. (b) Magnetic field dependence of $\chi''$ and $\chi'$
         at $\omega/2\pi = 1065$ Hz in high magnetic fields.
Insets: (a) $\chi''(h_0)$ dependence at $T=36$~K; (b)
$\chi_{dc}\equiv M(H_0)/H_0$ for the virgin curve, and $\chi'(H_0)$
         in whole range of dc field.}
     \label{f-1}
     \end{center}
     \end{figure}

Assuming that in zero dc field there is a complete screening of ac
field without any losses, one could measure the absolute value of ac
susceptibility, $\chi'$ and $\chi''$, as a function of external
parameters, such as dc field, frequency, and temperature. At low ac
amplitudes the response is linear and does not depend on $h_0$ as
shown in the inset~(a) to Fig.~\ref{f-1}b, where the $\chi''$ at
$\omega/2\pi=1000$~Hz, $H_0=4.3$~kOe, and $T= 36$~K is plotted as a
function of $h_0$. Fig.~\ref{f-1}a shows magnetization loop $M(H_0)$
at $T = 34$~K (upper inset) and the high field $M(H_0)$ data in an
extended scale (main panel), where the irreversibility field,
$H_{irr}$ and $H_{c2}$ are indicated. The lower inset of
Fig.~\ref{f-1}a presents $H_{irr}(T)$ curve. These results are
consistent with the data reported previously (see, for
example,~\cite{CAN} and references therein).

\begin{figure}
     \begin{center}
    \leavevmode
 \includegraphics[width=0.9\linewidth]{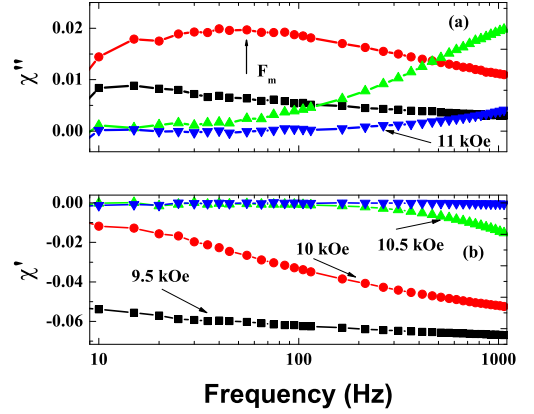}

\caption{(Color online) Frequency dependence of $\chi''$ (a) and
$\chi'$ (b) at $T=34$~K for different magnetic fields. The symbols
for the fields in both figures are identical.}
     \label{f-2}
     \end{center}
     \end{figure}

 The inset (b) in Fig.~\ref{f-1}b shows dc and ac susceptibility measured at 34 K as a
function of dc field. It is evident that the $\chi_{dc}\equiv
M(H_0)/H_0$ is already very small at 5 kOe, whereas the $\chi'$
becomes zero only above 12 kOe. The dc field dependence of $\chi'$
and of $\chi''$ is shown in Fig.~\ref{f-1}b (main panel). Here
again, the $\chi'$ curve exhibits ideal diamagnetism in fields much
higher than 5 kOe. Note the high peak of losses at about 11 kOe,
Fig.~\ref{f-1}b, which considerably exceeds the losses both in the
normal and Meissner states. A frequency dispersion of $\chi$ is also
evident (see Fig.~\ref{f-2}). All these features are typical for the
ac response in SSS of low-temperature superconductors (LTS)
~\cite{ROLL,GLT}. However, there is an apparent difference. In
isotropic LTS these features are observed only at $H_0>H_{c2}$,
where bulk dc magnetization is zero. In contrast, in polycrystalline
MgB$_2$ SSS are observed below $H_{c2}$ (see Figs.~\ref{f-1}a
and~\ref{f-1}b) and coexist with dc bulk magnetization. The dc
moment in this region is reversible and small, but unambiguously has
a bulk origin. In order to prove this, we applied a low-frequency
shaking field with the amplitude of 2.5 Oe. We did not observe any
change of the dc magnetization curve. Any possible nonzero surface
current contribution to dc magnetic moment has a nonequilibrium
character if the volume is in the normal state, and one could expect
hysteresis and disappearance of the dc magnetization after the
application of the shaking ac field. On the other hand, the ac
response cannot be attributed to bulk properties. For example, at
$T= 34$~K and $H_0=10.2$~kOe  $\chi'$ at 1065~Hz is about 0.5 (the
midpoint of the transition in Fig.~\ref{f-1}b, main panel). This
value is 3 orders of magnitude higher than the experimental $
dM/dH_0$ measured under the same conditions.

\begin{figure}
     \begin{center}
    \leavevmode
 \includegraphics[width=0.9\linewidth]{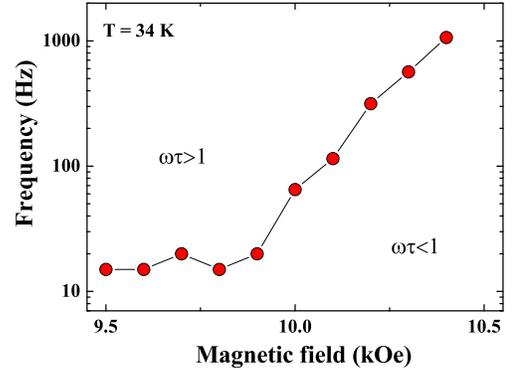}

\caption{(Color online) Magnetic field dependence of $F_m$.}
     \label{f-3}
     \end{center}
     \end{figure}

The frequency dependence of $\chi'$ and $\chi''$ at constant dc
fields is shown in Fig.~\ref{f-2}. It is clear that in some fields
$\chi'(\omega)\propto\ln(\omega)$. $\chi''$ show a maximum at
$\omega/2\pi=F_m$ and this maximum moves with dc field as shown in
Fig.~\ref{f-3}. One may predict that, similar to the spin-glass
state, the low frequency dynamics of SSS could be characterized by a
broad spectrum of relaxation times with some maximum value $\tau$.
Indeed, Fig.~\ref{f-3} shows that $F_m(H_0)$ divides the
\emph{frequency - magnetic field } phase diagram into two parts,
above and below the line, where $\omega\tau>1$ and $\omega\tau<1$,
respectively.

GL equations for two-band anisotropic superconductors were discussed in numbers of
articles~\cite{GUR,ASK,ZT} and in linear approximation these take the form:

\begin{equation}\label{Eq1}
\mu_{sij}\Pi_i\Pi_j\Delta_s-\alpha_s(T)\Delta_s-\delta_{ss'}\Delta_{s'}=0
\end{equation}
where the index $s=1$(2) corresponds to the $\sigma$($\pi$) bands, $\Pi_j\equiv
i\nabla_j+2eA_j/\hbar c$, $\overrightarrow{A}$ - vector potential, $\Delta_s$ - order
parameter of $s$ band, $\widehat{\mu}_s$ - inverse mass tensor and $ \delta_{ss'}$ is
different from zero only if $s\neq s'$. The coefficients in these equations,
$\widehat{\mu}_s$, $\alpha_s(T)$ and $\delta_{ss'}$, are given through the BCS
superconducting coupling matrix $\lambda_{nm}$ and could be found in~\cite{GUR,ZT}. Let
us consider the superconducting slab of thickness $2D$ in a magnetic field parallel to
its surface and chose the coordinate system with $x$-axis perpendicular to the surface,
magnetic field along $z$-axis and the plane $x=0$ in the center of the slab. The second,
$H_{c2}$, and the third critical magnetic fields, $H_{c3}$, are determined as the maximum
fields for which the eigen solutions of linearized Eq.~(\ref{Eq1}) with vector potential
$\overrightarrow{A}(0,Hx,0)$ and appropriate boundary conditions could be found. Looking
for the solution of Eq.~(\ref{Eq1}) in the form:

\begin{equation}\label{Eq2}
\Delta_s=f_s(x)\exp(iky+ik_zz-i\widetilde{\mu}_{sxz}k_zx+i\widetilde{\mu}_{sxy}hx^2/2)
\end{equation}
one obtains

\begin{equation}\label{Eq4}
\begin{array}{c}
\frac{d^2f_s}{dx^2}-(\widetilde{\mu}_{syy}-\widetilde{\mu}_{sxy}^2)h^2(x-x_s)^2f_s+
(\widetilde{\alpha} +g)f_s+\\
\widetilde{\delta}_{ss'}f_{s'}exp(i\phi_{s'})=0\\
g=\frac{(\widetilde{\mu}_{yz}-\widetilde{\mu}_{xy}\widetilde{\mu}_{xz})^2-
(\widetilde{\mu}_{zz}-\widetilde{\mu}^2_{xz})(\widetilde{\mu}_{yy}-\widetilde{\mu}_{xy}^2)}
{\widetilde{\mu}_{yy}-\widetilde{\mu}^2_{xy}}<0\\
x_s=k/h+\frac{\widetilde{\mu}_{yz}-\widetilde{\mu}_{xy}\widetilde{\mu}_{xz}}
{\widetilde{\mu}_{yy}-\widetilde{\mu}^2_{xy}}k_z/h,
\end{array}
\end{equation}
with $\phi_1=-\phi_2=(\widetilde{\mu}_{1xy}-\widetilde{\mu}_{2xy})(kx-hx^2/2)+
(\widetilde{\mu}_{1xz}-\widetilde{\mu}_{2xz})k_zx$, and boundary conditions $
df_s(D)/dx=0$. Here the dimensionless variables are used: $h=2eH_0/\hbar c$,
$\widetilde{\mu}_{sik}=\mu_{sik}/\mu_{sxx}$, $\widetilde{\alpha}_{s}=\alpha_s/\mu_{sxx}$,
$\widetilde {\delta}_{ss'}=\delta_{ss'}/\mu_{sxx}$. $H_{c2}$ is determined by the maximum
field for which the solution of Eqs.~(\ref{Eq4}) exists at $k=k_z=0$ with boundary
conditions $df_s(0)/dx=0$ and $df_s(D)/dx=0$ and $H_{c3}$ is determined by the solutions
of this equation with $k\approx hD$, $f_s(0)=0$ and $df_s(D)/dx=0$.

\begin{figure}
     \begin{center}
    \leavevmode
 \includegraphics[width=0.9\linewidth]{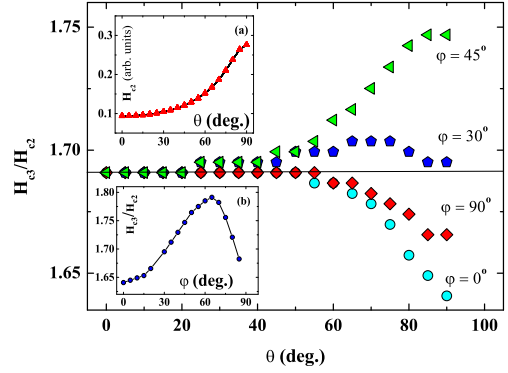}

 \caption{(Color online) The ratio $H_{c3}/H_{c2}$
 as a function of polar angle $\theta$ for $\varphi = 0$, 30, 45, and 90 degree.
       Insets: (a) angular dependence of $H_{c2}(\theta)$; (b) $H_{c3}/H_{c2}(\varphi)$
       for $\theta=90^{\circ}$.}
     \label{f-4}
     \end{center}
     \end{figure}

At first we discuss the one-band anisotropic superconductor,
$\delta_{ss'}=0$. Equations~(\ref{Eq4}) are of the same type as in
the isotropic case with GL parameter
$\kappa=\sqrt{\widetilde{\mu}_{yy}-\widetilde{\mu}^2_{xy}}$, and
using the known solution~\cite{PG} we obtain
$H_{c3}=1.695\alpha(\mu_{xx}\mu_{yy}-\mu^2_{xy})^{-1/2}$. $H_{c2}$
is also determined by Eq.~(\ref{Eq4}) and equals to
$\alpha(\mu_{xx}\mu_{yy}-\mu^2_{xy})^{-1/2}$~\cite{KOG}. Thus the
ratio $H_{c3}/H_{c2}$ for anisotropic crystals does not depend on
the orientations of principal axes and is the same as for the
isotropic case. For a two-band superconductor with different
$\mu_{ik}$ the analytical solution hardly could be found and the
numerical simulations can be successfully used. For an uniaxial
crystal, $H_{c2}$  depends only on the angle $\theta$ between the
$c$-axis and the direction of the dc magnetic field. In
polycrystalline samples only the grains with $H_{c2}>H_0$ give a
contribution into the equilibrium magnetic moment~\cite{CLM,BUD}:

\begin{equation}\label{Eq5}
\begin{array}{c}
M_{eq}(H_0)=\int^{\pi/2}_{0}\frac{(H_0-H_{c2}(\theta))}{8\pi\beta_c\kappa^2}\times\\
\Theta(H_{c2}(\theta)-H_0) \varepsilon(\theta)sin(\theta)d\theta,
\end{array}
\end{equation}
where function $\varepsilon(\theta)$ also takes into account the
anisotropy of the vortex lattice and is equal to 1 for isotropic
sample, $\beta_c=1.16$, and $\Theta(y)=1$ for $y>0$ and zero
otherwise. For an uniaxial two-band superconductor,
$\varepsilon(\theta)$ is not known yet while for the one-band
superconductor it was calculated in~\cite{CLM},
$\varepsilon(\theta)=[\sin^2(\theta)+\gamma^2\cos^2(\theta)]/\gamma^{2/3}$,
where the anisotropy parameter $\gamma=H_{c2}(\pi/2)/H_{c2}(0)$.
Using for the coupling matrix from ~\cite{RYD} $\lambda_{11}=0.81$,
$\lambda_{22}=0.285$, $\lambda_{12}=0.119$, $\lambda_{21}=0.091$ and
$\mu_{2c}/\mu_{2a}=1.487$, $\mu_{1a}/\mu_{2a}=0.2$~\cite{ZT} and
varying $\mu_{1c}/\mu_{1a}$, the equilibrium magnetization near
$H_{c2}$ have been calculated for different temperatures. We found
that using $\mu_{ik}$ for clean limit with $\mu_{1c}/\mu_{1a}=0.048$
and $\varepsilon =1$ gives the best fit to experimental data. The
result of these calculations is shown in Fig.~\ref{f-1}a  by
triangles. The GL parameters obtained from this procedure are:
$\kappa=18.6$, 16.1 and 13.8 for $T= 34$~K, 36~K and 37~K,
respectively. The GL formulae for one band isotropic superconductor
with $\kappa=H_{c2}/H_c\sqrt 2$, where $H_c$ is the thermodynamic
critical field and $M_{eq}=(H_0-H_{c2})/8\pi\beta_c\kappa^2$, gives
$\kappa=7.7$, 7.8, and 13.9 for the same temperatures. At $T= 36$~K
and $H_0=3$~kOe, $M_{eq}$ from this formula is 1.81~emu/cm$^3$,
whereas the experimental value is significantly smaller,
0.11~emu/cm$^3$. The calculated $H_{c3}/H_{c2}$ ratio for a single
crystal equals, with accuracy $\approx 10\%$, a value of 1.69
predicted in~\cite{PG} for all orientations of principal axes,
whereas both $H_{c2}$ and $H_{c3}$ strongly depend on orientation.
Fig.~\ref{f-4} shows $H_{c3}/H_{c2}$ as a function of the polar,
$\theta$, and azimuthal, $\varphi$, angles using the same set of
material parameters as above for $T=34$~K that corresponds to
$T/T_c\approx 0.9$. The inset~(a) in Fig.~\ref{f-4} shows $H_{c2}$
as a function of $\theta$. The ratio $H_{c3}/H_{c2}(\varphi)$ is
shown in the inset (b) in Fig.~\ref{f-4}. The calculated anisotropy
parameter $\gamma=2.9$ is in quantitative agreement with $\gamma$
reported in~\cite{ZEH} and somewhat larger than found in~\cite{WLP}.

The sample investigated in the present work is composed of randomly
oriented grains with average size of about 40 nm and evenly
distributed $H_{c3}$. The observed complete screening of ac field
shows that the whole sample surface is in SSS. The onset of the
screening could be considered as the percolation transition and the
appearance of the large continuous clusters in SSS. Percolation
transition in SSS of Nb was discussed in Ref.~\cite{JUR}.
\begin{figure}
     \begin{center}
    \leavevmode
 \includegraphics[width=0.9\linewidth]{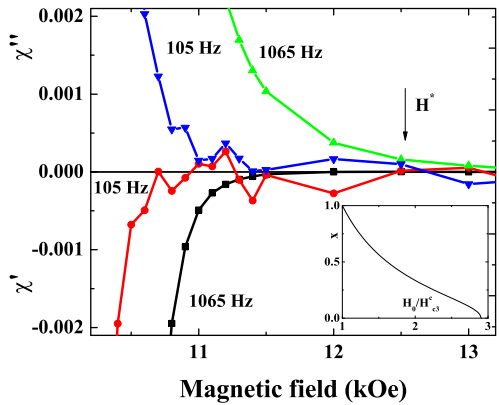}

       \caption{(Color dependence) Magnetic field dependence of $\chi'$ and $\chi''$ at
       $T=34$~K in high magnetic fields. $H^*$ is shown for
       $\omega/2\pi=1065$~Hz.
        The inset shows fraction of grains in SSS, $x$, as a function of $H_0/H^c_{c3}$.}
     \label{f-5}
     \end{center}
     \end{figure}
Fig.~\ref{f-5} shows the expanded view of $\chi'$ and $\chi''$ at
large fields and $T=34$~K at two frequencies $\omega/2\pi =105$ and
1065 Hz. The physical nature of the displacement of the curves along
the $H$-axis with frequency is not clear yet, but the response at
1065 Hz gives a more accurate value of the field $H^*$, at which the
percolation transition takes place. The $H^*$ found from $\chi'$
equals 11.5~kOe. The $H_{c2}$ at $T=34$~K is about 14.5 kOe. Using
the calculated above $\gamma$=2.9, one could find that parallel to
the $c$-axis $H_{c3}^{c}$ is about 8.5 kOe and
$H^*/H_{c3}^{c}\approx 1.35$. Assuming random uniform distribution
of $c$-axis orientation in the grains and
$H_{c3}(\theta\varphi)/H_{c2}(\theta)\cong 1.7$ we calculated the
fraction of the grains in SSS on the surface as a function of dc
magnetic field,
$x(H_0)=\int_{0}^{\pi/2}\sin(\theta)\Theta(1.7H_{c2}-H_0)d\theta$.
The result of this calculation is shown in the inset to
Fig.~\ref{f-5}. The value $H^*/H_{c3}^{c}=1.35$ corresponds to the
fraction of superconducting grains $x\approx 0.66$. One could infer
that $\chi''$ provides a more accurate onset for SSS, then
$H^*=12.5$~kOe and $x\approx 0.58$. These values are in good
agreement with the result for critical concentration of conducting
clusters $x_c=0.6$ in the $2d$ percolation site problem~\cite{GANT}.
We consider this as an additional argument in favor of the surface
nature of the observed ac response.

The adequateness of the GL model for the description of MgB$_2$ was discussed recently in
~\cite{KG} in frame of the Usadel equations. In the dirty limit, the GL model is correct
if the dimensionless parameter $\eta=q^2D_c/2\pi T$ is less than 1. Here $q\approx 1/x_0$
is the average Fourier component of $f_s(x)$ in Eq.~(\ref{Eq2}) and $D_c$ is the
diffusion coefficient through which the tensor $\widehat{\mu}$ could be
expressed~\cite{GUR}. To be sure that calculations of $H_{c2}$ and $H_{c3}$ are correct
we found parameter $\eta$ for the most sensitive orientation of $c$-axis,
$\overrightarrow{H_0}\perp c$ and $\varphi=\pi/2$ assuming that $x_0$ is the distance at
which the order parameter in the $\pi$-band decreases by the factor of three. It was
found that $\eta<1$ for $T/T_c>0.8$ and GL equations could be used at $T/T_c>0.8$.
Another argument in favor of the GL equations is the very good agreement between the
temperature dependence of $H_{c3}/H_{c2}$ for $H_0$ in $ab$ plane calculated from
Eq.~(\ref{Eq1}) and the result obtained in~\cite{DG} on the basis of Usadel model.

In summary, we have demonstrated the existence of SSS in
polycrystalline MgB$_2$. The SSS coexist with weak bulk
magnetization. The ac response is linear at low amplitudes of
excitation field. The transition to SSS is of $2d$ percolation
character. The frequency dispersion for $\omega/2\pi$ in the
interval of 5-1065~Hz shows a maximum in $\chi''$, and $\chi'$ is
proportional to $\ln\omega$. The superconducting current in the
grain depends on the orientation of the principal axes and is a
function of the instant values of magnetic field and both $k$ and
$k_z$. The relaxation of $k$ and $k_z$ to their equilibrium values
with zero surface current determines the ac response of the grain.
The subsequent averaging over all grains provides a complex
character of the observed ac response. We found that GL equations
adequately describe the equilibrium dc magnetization of MgB$_2$
samples at large field and could be used for analysis of SSS. It was
shown that the ratio $H_{c3}/H_{c2}$ weakly depends on the
orientation of the $c$-axis whereas both $H_{c2}$ and $H_{c3}$
strongly depend on the orientation, which leads to the coexistence
of SSS and weak bulk dc magnetic response.

This work was supported by the Klatchky foundation for
superconductivity. We wish to thank E.B. Sonin for many helpful
discussions.

\end{document}